\documentclass[floatfix,10pt,aps,prb,twocolumn,showpacs,a4paper]{revtex4}

\usepackage{amsmath}   
\usepackage{mathtools} 
\usepackage{amssymb}   
\usepackage{mathrsfs}  
\usepackage{textcomp}  
\usepackage{accents}   
\usepackage{bm}        
\usepackage{soul}
\usepackage{relsize}
\usepackage{graphicx}  
\usepackage{epstopdf}  
\usepackage{subfigure} 
\usepackage{makecell}
\usepackage{hhline}
\usepackage{cellspace}
\usepackage{array}     
\usepackage{tabularx}  
\usepackage{multirow}  
\usepackage{booktabs}  
\usepackage{dcolumn}   
\usepackage{gensymb}   
\usepackage{setspace}
\usepackage{scrextend}
\usepackage[usenames,dvipsnames]{color} 
\usepackage{url}       
\usepackage{hyperref}  
\hypersetup{colorlinks,citecolor=Violet,linkcolor=Red,urlcolor=Blue}
\newcommand{\FeFThree}{FeF$_{3}$}
\graphicspath{{./}}
\begin{document}
\setlength{\heavyrulewidth}{0.08em}
\setlength{\lightrulewidth}{0.05em}
\setlength{\cmidrulewidth}{0.03em}
\setlength{\belowrulesep}{0.65ex}
\setlength{\belowbottomsep}{0.00pt}
\setlength{\aboverulesep}{0.40ex}
\setlength{\abovetopsep}{0.00pt}
\setlength{\cmidrulesep}{\doublerulesep}
\setlength{\cmidrulekern}{0.50em}
\setlength{\defaultaddspace}{0.50em}
\setlength{\tabcolsep}{4pt}
\title{Crossover between tricritical and Lifshitz points in pyrochlore \FeFThree{} }

\author{Mohammad Amirabbasi}
\affiliation{Department of Physics, Isfahan University of Technology (IUT), Isfahan 84156-83111, Iran}
\author{Nafiseh Rezaei}
\affiliation{Department of Physics, Isfahan University of Technology (IUT), Isfahan 84156-83111, Iran}
\author{Mojtaba Alaei}
\email {m.alaei@cc.iut.ac.ir}
\affiliation{Department of Physics, Isfahan University of Technology (IUT), Isfahan 84156-83111, Iran}
\author{Farhad Shahbazi}
\affiliation{Department of Physics, Isfahan University of Technology (IUT), Isfahan 84156-83111, Iran}
\author{Hadi Akbarzadeh}
\affiliation{Department of Physics, Isfahan University of Technology (IUT), Isfahan 84156-83111, Iran}
\date{\today}
\begin{abstract}
Pyrochlore FeF$_{3}$ (pyr-FeF$_{3}$) is a  Heisenberg anti-ferromagnetic (AF) with a magnetic susceptibility deviating  from the Curie-Weiss law, even at the room temperature. This compound shows a transition to a long-range ordered state with all-in all-out (AIAO) spin configuration. The critical properties of  this transition have  remained a matter of dispute.   
In this work, to gain more insight into the critical properties of pyr-FeF$_{3}$, using ab initio density functional theory (DFT), we obtain  spin Hamiltonian of this material under the relative volume change with respect to the experimental volume ($\frac{\Delta V}{V_0}$) from $-0.2$ to $0.2$. We show that the relevant terms in the spin Hamiltonians are the AF exchange up to third neighbors, the nearest neighbor bi-quadratic and  the direct Dyzaloshinski-Moriya (DM) interactions and find how these  coupling constants vary under  the volume change.  Then we study the effect of volume change on the finite temperature critical behavior, using classical Monte Carlo (MC) simulation. 
We show that the spin system undergoes a weakly first order transition to AIAO at small volumes which turns to  a second order transition close to the experimental structure. However, increasing $\frac{\Delta V}{V_0}$ to $\sim0.2$, systems shows  a transition to a modular  spin structure. This finding suggests the existence of a Lifshitz point in  pyr-FeF$_{3}$ and may explain the unusual critical exponents observed for this compound.    
\end{abstract}

\maketitle

\section{introduction}
\label{sec:introduction}

During the past few years, curious behavior of geometric frustrated pyrochlores have been a topic of constant significance due to their interesting peculiar physics \cite{Zhou2017,Petit2016,Ramirez1994,Henelius2016,Kapellasite2012,Gardner2010,Sohn2017,Zhang2017,Ueda2012,sadeghi2015,Lacroix}.
Geometrically frustrated pyrochlore is a three-dimensional lattice consisting of corner-sharing  tetrahedra in which the magnetic ions are placed on the corner of each tetrahedron, mostly with anti-ferromagnetic interaction between nearest neighbors. 
In these systems, magnetic properties deviate from conventional magnetic systems, in a sense that magnetic moments do not tend  to form a long range ordering even at the temperatures much below the Curie-Weiss temperature ($\Theta_\mathrm{CW}$). The reason for such a strange behavior is in the geometry of the system where the local energy optimization does not tend to a unique global minimum,  which gives rise to an extensive ground state degeneracy.  As a result, geometrically frustrated materials have been found to exhibit a wealth of exotic ground states and behaviors such as spin glass\cite{Singh2012}, spin ice \cite{Bramwell2001} or even spin liquid\cite{Savary2017}.
pyr-FeF$_{3}$ is an  anti-ferromagnet Heisenberg pyrochlore which shows a transition to a long range ordered state at a temperature much smaller than its Curie-Weiss temperature and its susceptibility also deviates from the Curie-Weiss law even at room temperature \cite{Ferey1986}.
The structure of pyr-FeF$_{3}$ is faced-center cubic  with space group Fd\={3}m, where Fe atoms occupy the 16c (0,0,0) sites and fluorine the 48f (x, 1/8, 1/8) sites \cite{Pape1986}. Indeed, the Fe atoms reside on the corners of each tetrahedron while each fluorine  lays in a position between any two irons  but not exactly on the edge, henceforth giving rise to a 142.3$^\circ$ Fe-F-Fe bond angle. 
The experimental values for the lattice parameter and internal parameter (x) (which determines the Fe-F-Fe bond angle) are 19.511 (a.u.) and 0.3104(5), respectively\cite{Pape1986}. 
Experimental results\cite{Calage1987} (M{\"o}ssbauer experiments) show that the transition temperature is about 20 K where, below this temperature, the Fe magnetic moments  point toward or out of  the tetrahedron centers, that is the so called all-in/all-out (AIAO) ordering. 
Another interesting peculiarity of  pyr-FeF$_{3}$ is the universality class of  its transition to ALAO, measured by the order parameter exponent $\beta\sim 0.18\pm 0.02$~\cite{Reimerz1992} which deviates from the known universalities such as Ising ($\beta\approx 1/3$), Heisenberg ($\beta \approx 1/3$) or the tricritical point ($\beta=1/4$).  
In Ref. [\onlinecite{sadeghi2015}], using ab initio method based on density functional theory (DFT), we proposed a spin Hamiltonian for pyr-FeF$_{3}$ in its experimental structure. The dominant terms in the spin Hamiltonian were found to be the nearest neighbor AF Heisenberg, positive bi-quadratic   and direct   Dyzaloshinskii-Moriya (DM) interactions. The classical Monte Carlo (MC) simulation of this Hamiltonian reveals a transition to the ALAO state at the critical temperature $T_C=20$K and the value of the order parameter critical exponent was evaluated as $\beta=0.18\pm 0.02$ in agreement with the neutron scattering experiments\cite{Reimers1990}. It has also been shown that the systems enters in a Coulomb phase state for a wide temperature range from $20$K to $\sim100$K. 
The anomalous critical exponent observed for pyr-FeF$_{3}$, suggests that this compound in its experimental structure might be located near a multi-critical point. If so, a relatively little change in the parameters of the Hamiltonian may change the critical behavior of the system. Since, the magnetic exchange couplings are highly sensitive to the magnetic ion distances and the bond angles, the variation of  the spin Hamiltonian parameters can be calculated as a function of volume change in a way that all the symmetries of the lattice are preserved. 
Based on this motivation,  we aim to shed light on the critical properties of  pyr-FeF$_{3}$ by obtaining  its spin Hamiltonian in the presence of  hydrostatic positive (negative) pressure through decreasing (increasing) its volume at the ambient pressure.  Experimentally, the change in the volume can also be done through chemical pressure which means the substitution of some elements in a given compound  by smaller (positive pressure) or larger (negative pressure) elements ~\cite{Hallas2014}. Once the spin Hamiltonian is found, for each volume change the finite temperature critical behavior of the system can be investigated by the classical MC simulation.   

The paper is organized as the following. Section~\ref{sec:methodology} gives  the details of DFT method and MC simulation. In section~\ref{sec:results}, the structural variations under pressure, the resulting spin Hamiltonian for each volume change and the corresponding critical properties at finite temperature are discussed and finally, the end of this paper is devoted to the conclusion.  
\section{Computational methods}
\label{sec:methodology} 
In this paper, we employ Density Functional Theory (DFT) to construct an effective spin model Hamiltonian ($H_{spin}$) for pyr-FeF$_{3}$. 
The methods of calculation for different terms of spin Hamiltonian have been reported in~Ref.[\onlinecite{sadeghi2015}].
To this end, We use the FLEUR~\cite{fleur} (full-potential augmented plane wave (FLAPW) basis sets code).  
The exchange and correlation effects are considered using Perdew Burke Ernzerhof (PBE) functional from generalized gradient approximation (GGA)~\cite{Perdew1996}.
To improve electron-electron repulsion, we employ Hubbard correction within GGA+$U$ approximation.
For the anisotropic spin Hamiltonian term (the Dzyaloshinskii-Moriya (DM) and the single-ion interactions~\cite{sadeghi2015}), spin-orbit coupling effects are considered (GGA+$U$+SOC). Brillouin zone integrations are performed using $4\times4\times4$ k-points sampling for the conventional unit-cell (64 atoms) to calculate the Heisenberg exchange coupling constants and $6\times6\times6$ k-points sampling for primitive cell (16 atoms) for calculating the DM, single-ion  and bi-quadratic couplings.
We use 2.0 and 1.35 (a.u.) for the Muffin-tin radius of Fe and F atoms, respectively and optimized $k_{\mathrm{max}}=4.2\,(\mathrm{a.u.})^{-1}$ for cutoff energy.
Monte Carlo simulations are performed to find the critical properties of $H_{spin}$, using the  replica exchange method\cite{Hukushima1996}. 
We use three-dimensional lattices consisting of $N\times L^{3}$ spins, where $L$ is the linear size of the simulation cell, which in this work we take $L=4,5,6,7,8,9,10,11$.
For thermal equilibrium, we use $10^{6}$ Monte Carlo steps (MCS) per spin in each temperature and $10^{6}$ MCS for data collection. 
To reduce the correlation between the successive data, measurements are done after skipping 10 MCS.

\section{Results and Discussion}
\label{sec:results}
\subsection{Structural geometry under pressure}
It is known that in GGA+$U$ calculations, $U$ plays an important role, especially in magnetic properties. 
we  use  $U=3$eV and the Hund coupling $J_{H}=1$eV.  We show that choosing these values for the on-site Coulomb repulsion give rise to reasonable results for the magnetic properties of pyr-FeF$_{3}$ in its experimental structure at the  ambient pressure.  

\begin{figure}[htp]
    \centering
    \includegraphics[width=\columnwidth]{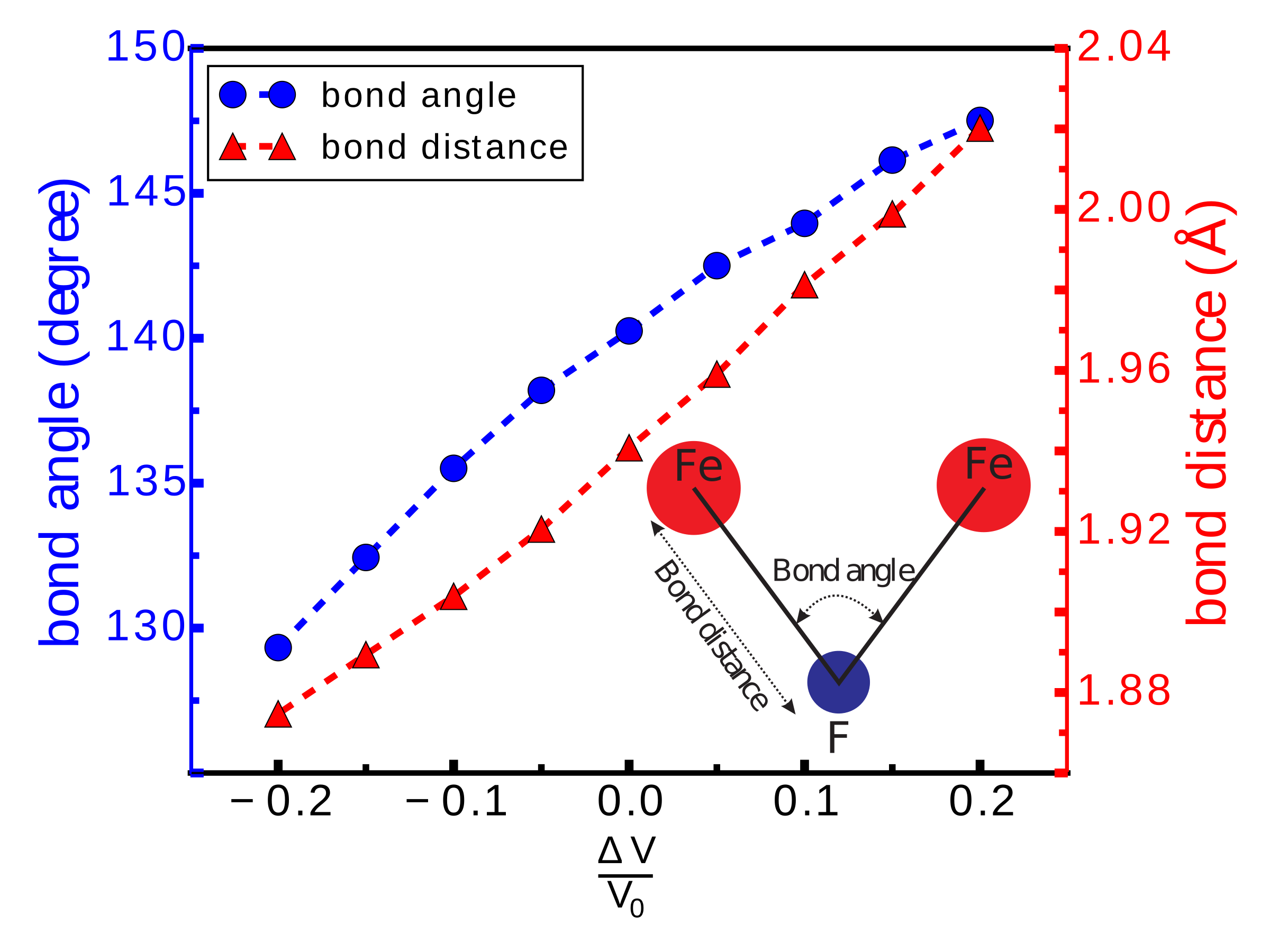} 
    \caption{(Color online) The GGA+$U$ result for the variations of the Fe-F-Fe bond angle and Fe-F bond distance in pyr-FeF$_{3}$ versus the fractional volume change $\frac{\Delta V}{V_0}$. $V_0$ denotes the unit cell volume of the experimental structure.}
    \label{fig:bond}
\end{figure}

In this section, we consider the effect of volume change on the structural geometry in GGA+$U$ calculation.
At each volume, we optimized the internal parameter ($x$) to find  the minimum total energy of  the system.
The results show that as the volume decreases the optimized $x$  also increases, meaning that the Fe atoms in the lattice get  closer to each other i.e. Fe-F-Fe bond angle and Fe-F bond distance become smaller. Fig.~\ref{fig:bond} shows the variations of  the bond angle (Fe-F-Fe) and the bond distance (Fe-F) with respect to the fractional volume change.
According to the Kanamori-Goodenough rule \cite{KANAMORI1959, Goodenough1955}, the larger bond angle (120-180) can lead to the anti-ferromagnetic exchange interaction between nearest neighbors, however, different bond distance has a more crucial effect on the strength of exchange interaction. 
So, at smaller lattice volumes, we expect to have stronger  exchange interaction due  to the shorter inter-ionic distance (see Fig.~\ref{fig:bond}).

\subsection{Spin Hamiltonian}
\begin{figure}[t]
    \centering
    \includegraphics[width=0.92\columnwidth]{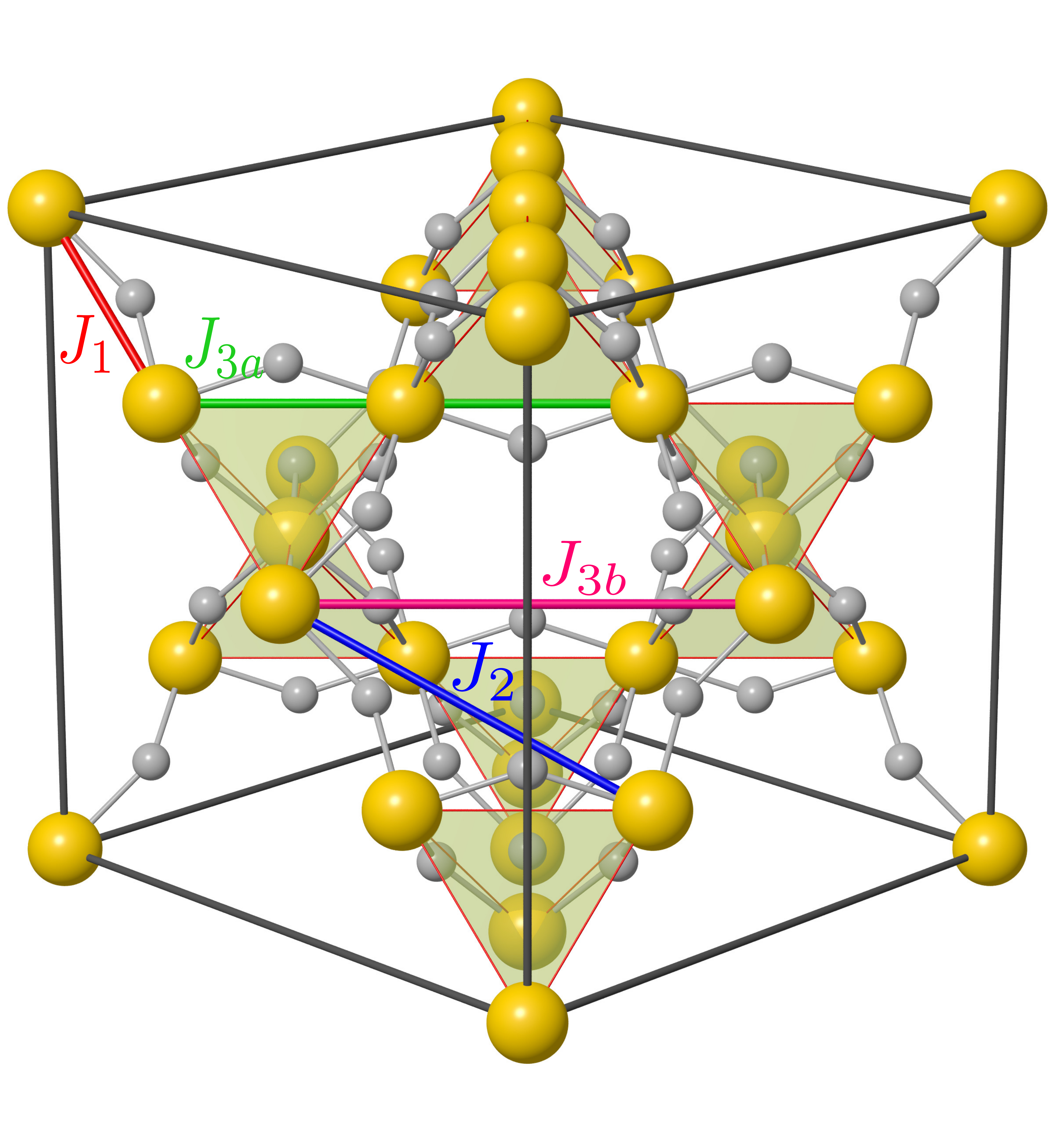}
    \caption{(Color online) pyr-FeF$_{3}$ structure with its corner-shared tetrahedrons in the cubic unit cell. Orange and gray spheres are $Fe^{3+}$ and $F^{-}$ ions, respectively. Different neighbors for Heisenberg interactions ($J_{1}, J_{2}, J_{3a}, J_{3b}$) have been shown.}
    \label{fig:J_parameter}
\end{figure}

\begin{figure}[b]
    \centering
    \includegraphics[width=0.92\columnwidth]{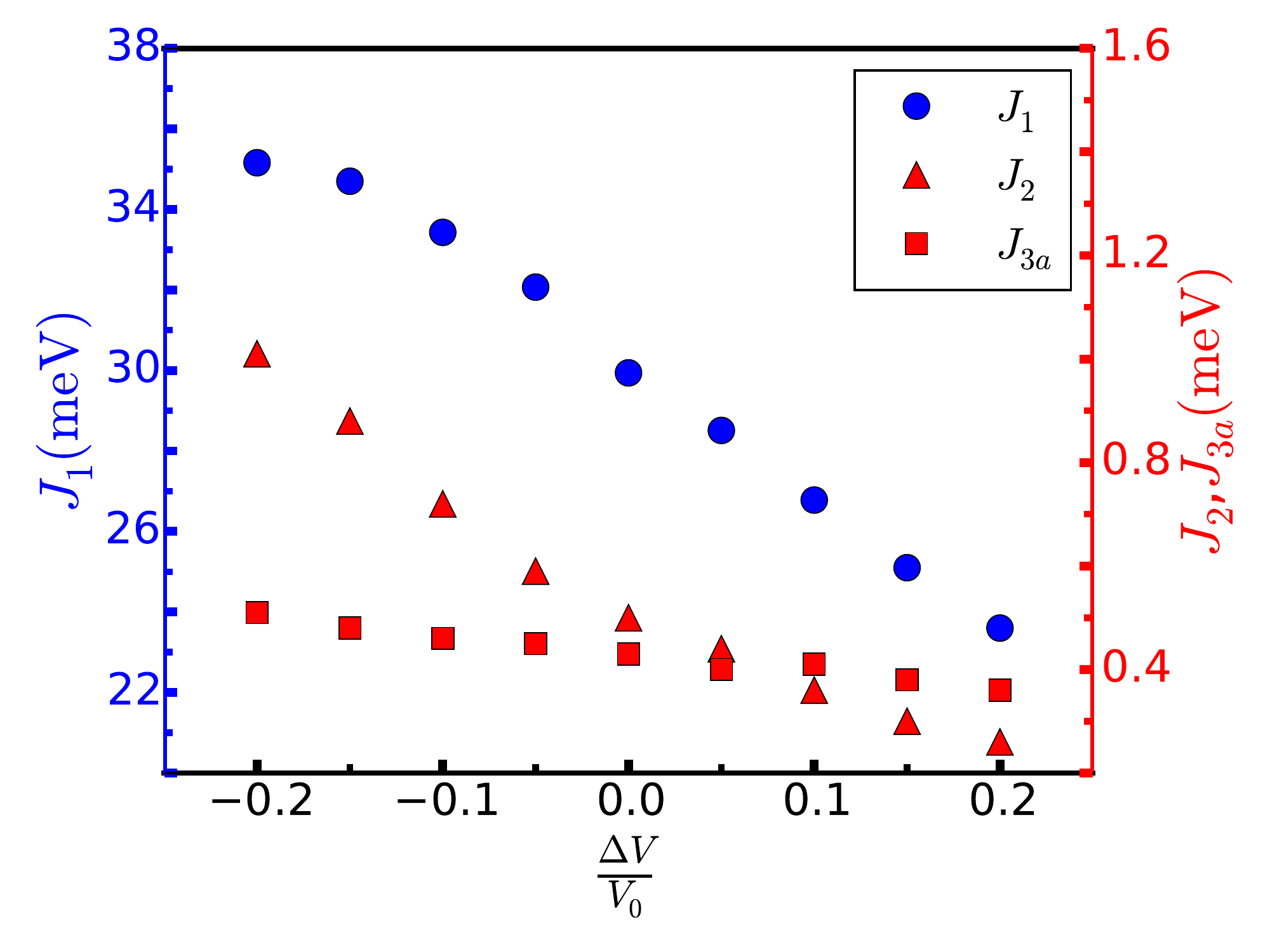} 
    \caption{(Color online) Heisenberg coupling constants $J_{1}, J_{2}$ and $J_{3a}$ versus  the fractional change in the unit cell volume, $\frac{\Delta V}{V_0}$. $V_0$ denotes the unit cell volume of the experimental structure. }
    \label{fig:Heisenberg-constants}
\end{figure}

Now we proceed to derive an effective Hamiltonian to find the ground state and also the finite temperature properties of pyr-FeF$_{3}$.
We define a model spin Hamiltonian that contains several spin-spin interactions, like Heisenberg, bi-quadratic, single-ion and direct Dzyaloshinskii-Moriya (DM) interactions:
\begin{equation}
\label{H}
\begin{split}
  H_{\rm {spin}} = \sum_{i\neq j} J_{ij}(\vec{S_{i}}\cdot\vec{S_{j}})+B\sum_{\rm n.n} (\vec{S_{i}}\cdot\vec{S_{j}})^{2} \\
+D \sum_{\rm n.n} \hat{D}_{ij}\cdot(\vec{S_{i}}\times \vec{S_{j}})+\Delta\sum_{i} (\vec{S_{i}}\cdot\vec{d_{i}})^{2}
\end{split}
\end{equation}
where $\vec{S_{i}}$ denotes a unit vector, $(J_{1}, J_{2}, J_{3a}, J_{3b})$ are the  Heisenberg couplings constants up to third neighbors (we assume $J_{3b}=0$, see Fig.~\ref{fig:J_parameter}), $B$ is the  bi-quadratic coupling constant between the nearest neighbors, 
$D$ and $\Delta$ denote the strengths of  DM and single-ion anisotropy, respectively. The unit vectors $\hat{D}_{ij}$ denote the directions of the direct DM vectors in pyrochlore \cite{Elhajal2005} .   
We estimate the Heisenberg coupling constants by mapping the collinear spin-polarized DFT total energies to   the spin Hamiltonian \eqref{H},
while for  the bi-quadratic, DM and single-ion parameters we used the total energies obtained by the non-collinear 
spin-polarized DFT.
Fig.~\ref{fig:Heisenberg-constants} shows  the variations of $J_{1}, J_{2}$ and $J_{3a}$ with respect to the fractional relative changes in the unit cell volume $\frac{\Delta V}{V_0}$, where $V_0$ denotes the unit cell volume of the experimental structure. This figure shows that the first and second neighbor Heisenberg coupling constants $J_1$ and $J_2$ rapidly decreases by increasing the volume, while the variation of $J_{3a}$ versus $\frac{\Delta V}{V_0}$ is much slower.     
\begin{figure}[t]
    \centering
    \includegraphics[width=0.98\columnwidth]{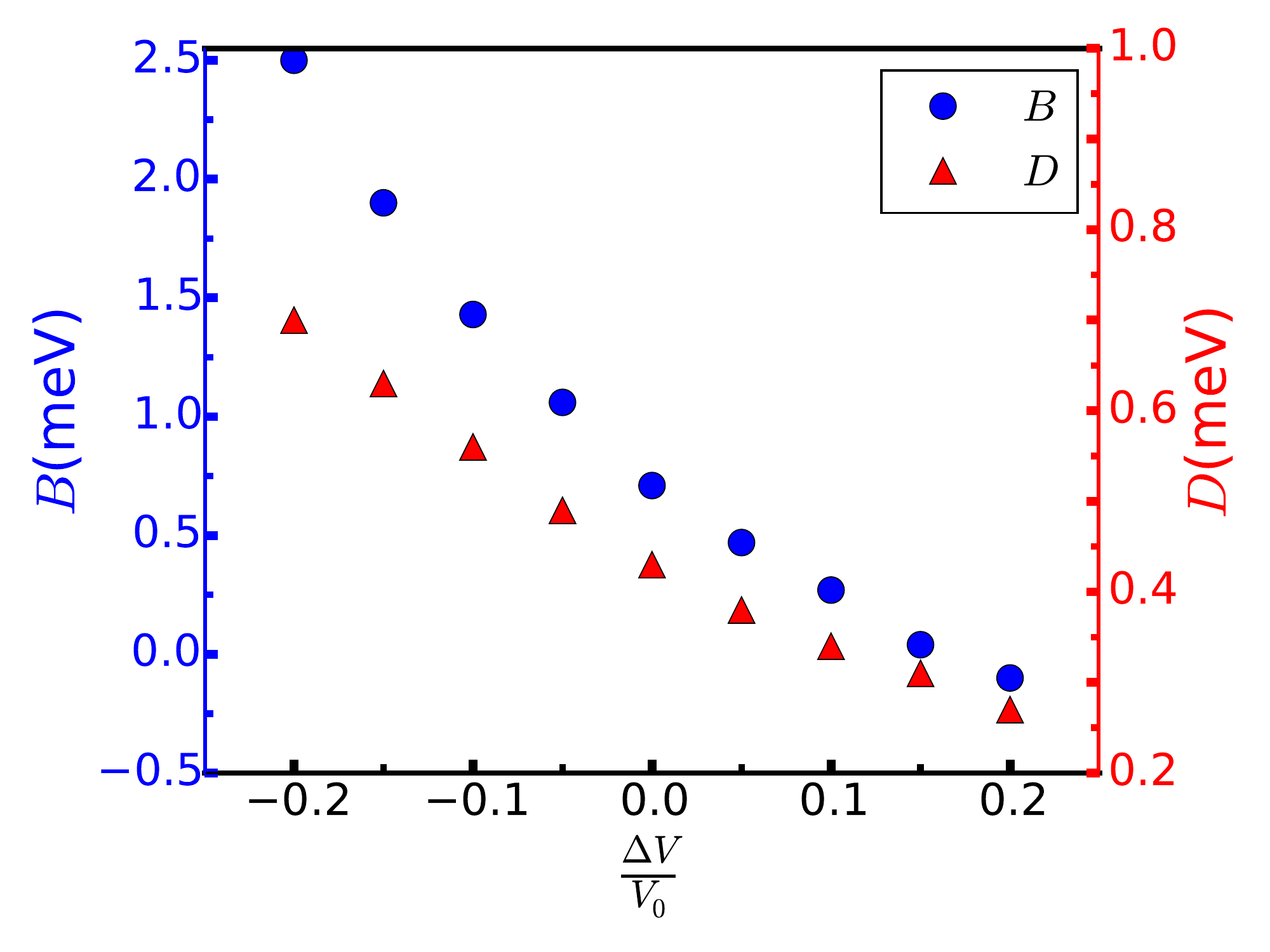} 
    \caption{(Color online) Variation of nearest neighbor bi-quadratic coupling $B$ and the DM coupling versus  the fractional change in the unit cell volume, $\frac{\Delta V}{V_0}$. $V_0$ denotes the unit cell volume of the experimental structure.}
    \label{fig:BD}
\end{figure}
\begin{figure}[b]
    \centering
    \includegraphics[width=0.92\columnwidth]{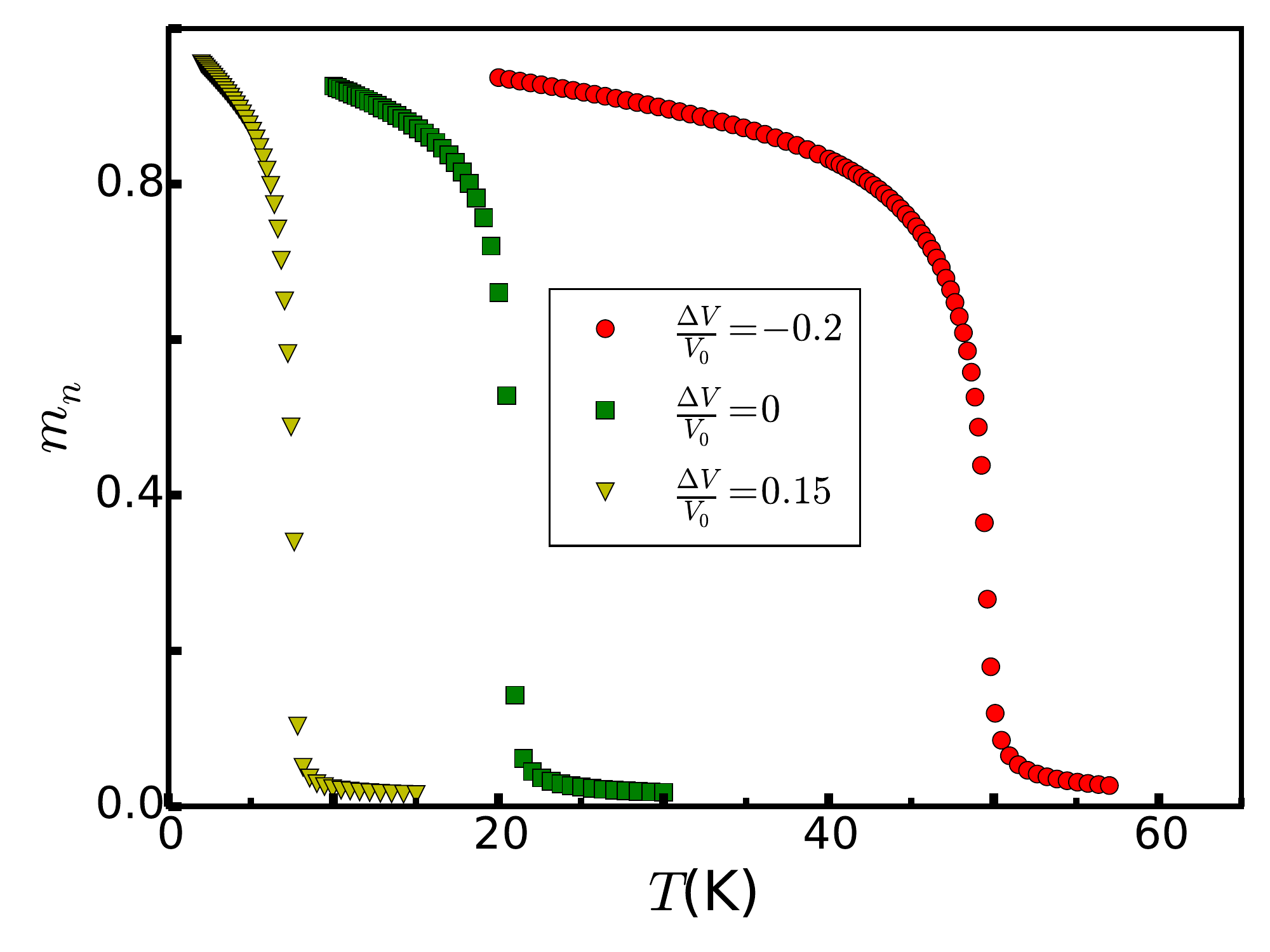} 
    \caption{(Color online) AIAO order parameter versus temperature for $\frac{\Delta V}{V_0}=-0.2, 0.0, 0.15$. The data are obtained by the MC simulations  on the  lattices with $L=10$.}
    \label{fig:m}
\end{figure}
Fig.~\ref{fig:BD} illustrates the  dependence of the $B$ and $D$ coupling constants on the fractional volumes changes, showing also the rapid fall of both interactions by increasing the unit cell volume.  Interestingly, these results show the sign change of the bi-quadratic coupling from positive to negative at $\frac{\Delta V}{V_0}\sim 0.15$.  
Our calculations for the strength of single ion anisotropy ($\Delta$) results that in all the unit cell volumes its value is an order of magnitude less than $B$ and $D$, so we can neglect this term. 


\subsection{Monte Carlo simulations}
\begin{figure}[t]
    \centering
    \includegraphics[width=0.92\columnwidth]{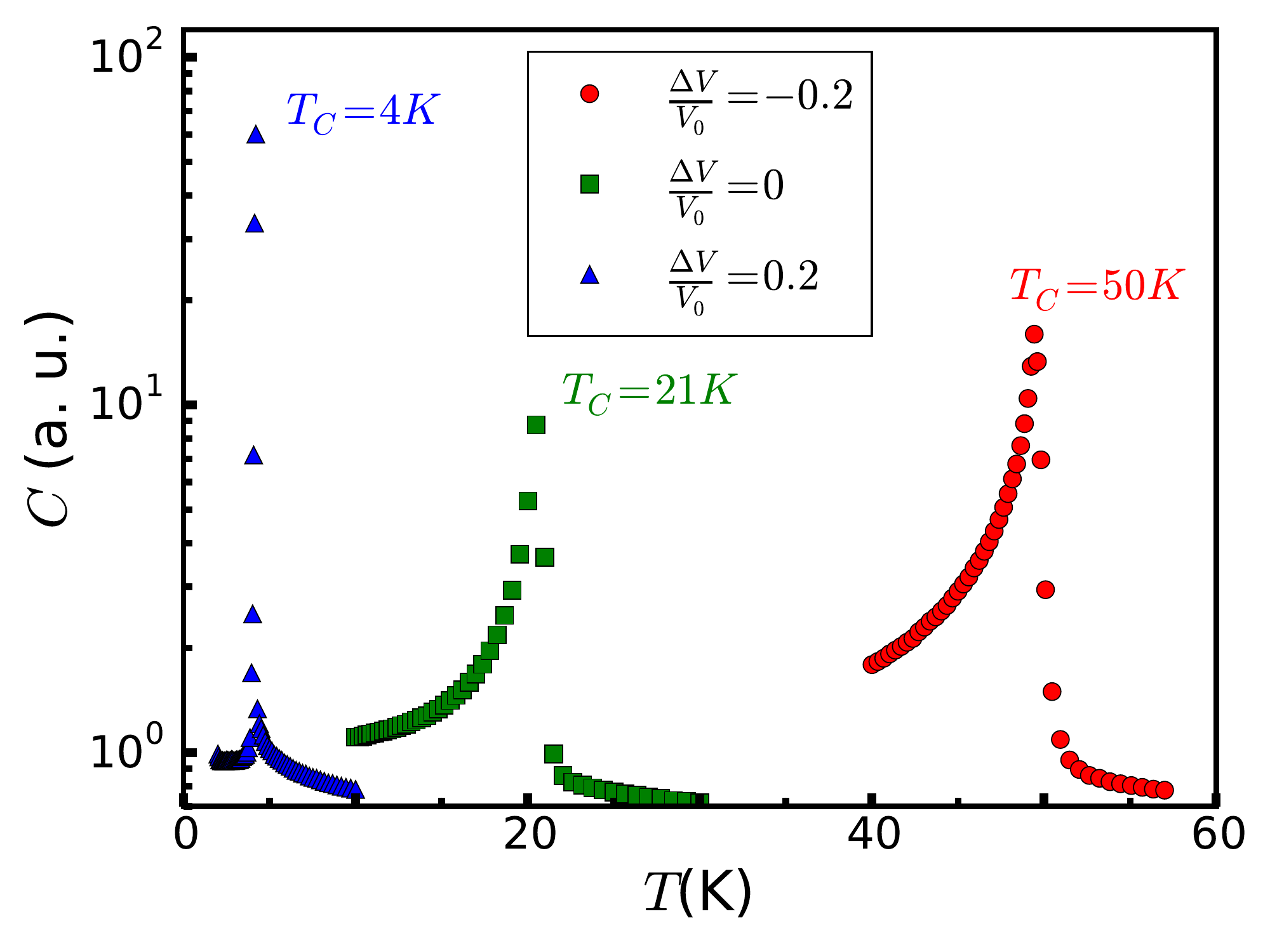} 
    \caption{(Color online) Specific heat versus temperature for $\frac{\Delta V}{V_0}=-0.2, 0.0, 0.2$. The data are obtained by the MC simulations  on the  lattices with $L=10$.}
    \label{fig:c}
\end{figure}

\begin{figure}[t]
    \centering
    \includegraphics[width=0.92\columnwidth]{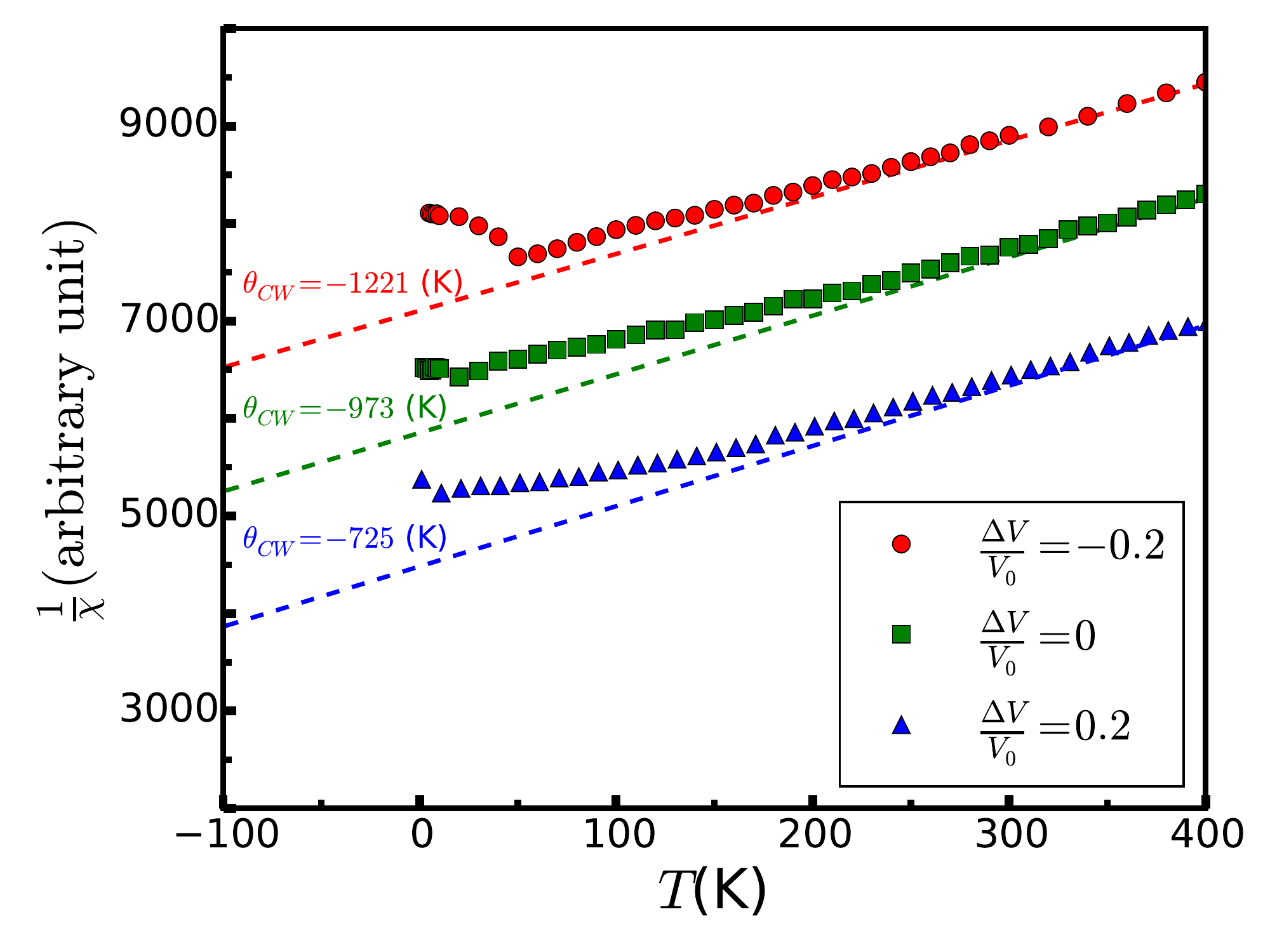} 
    \caption{(Color online) Temperature dependence of the inverse susceptibility for  $\frac{\Delta V}{V_0}=-0.2, 0.0, 0.2$. The data are obtained by the MC simulations  on the  lattices with $L=10$. The dashed lines represent the linear fits to the inverse susceptibility in the temperature range $300$K to $400$K, where the Curie law holds. The Curie-Weiss temperatures are obtained by the intersection of the dashed lines and  the T axis.   }
    \label{fig:kappa}
\end{figure}

\begin{figure}[b]
    \centering
     \includegraphics[width=0.92\columnwidth]{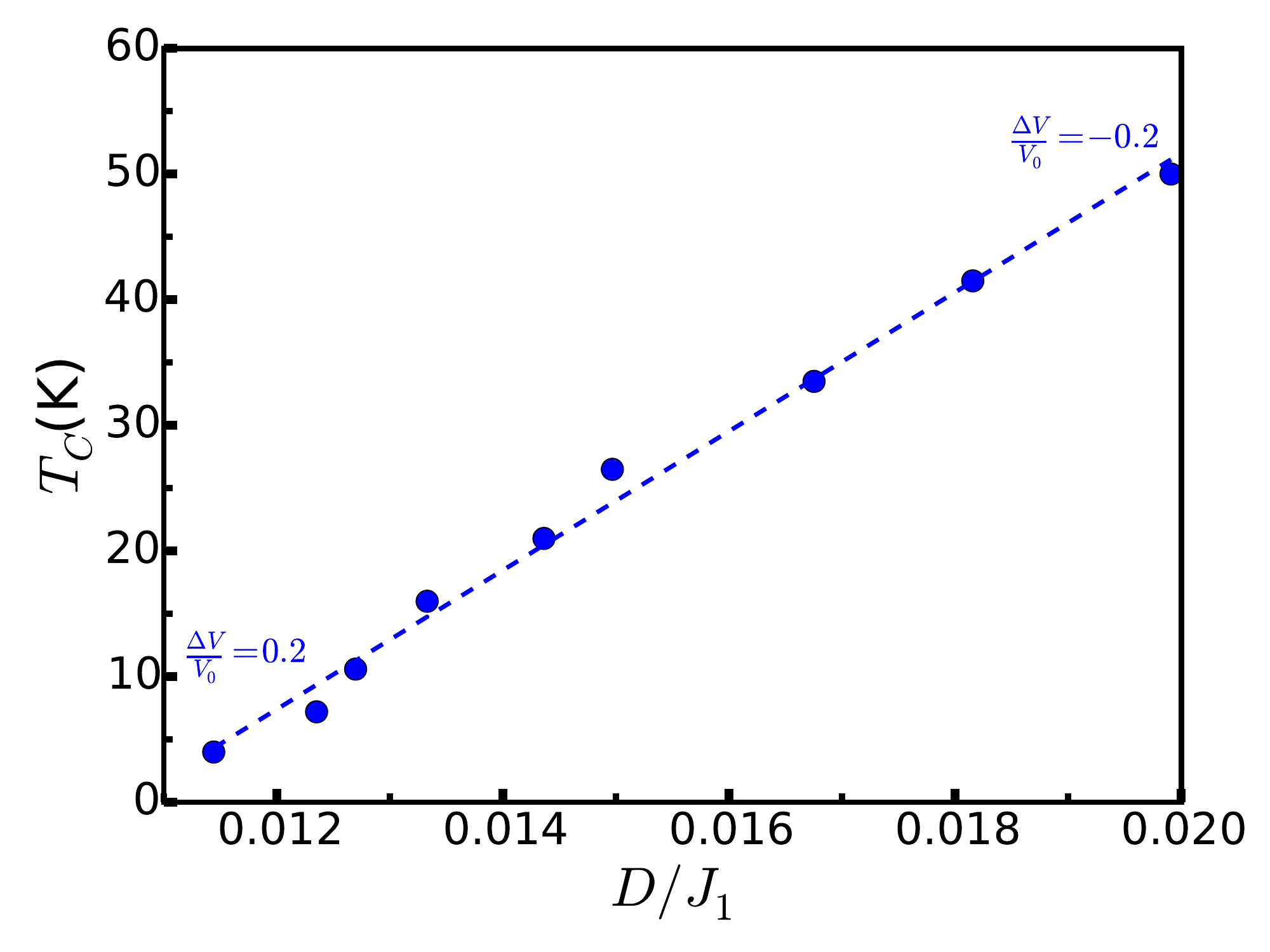}
     \caption{(Color online) Transition temperature $T_C$ versus $D/J_1$.}
    \label{fig:T_C}
 \end{figure}

In this section, we represent the MC results of the spin Hamiltonian \eqref{H} with the coupling constants obtained for the fractional volume changes $\frac{\Delta V}{V_0}=-0.2, -0.15, -0.1, -0.05, 0.0, 0.05 , 0.1, 0.15, 0.2$.  
We observe a phase transition to the all-in all-out (AIAO) long range order for $ -0.2\leq\frac{\Delta V}{V_0}\leq0.15$. This can be seen in Fig.~\ref{fig:m}, where the temperature behavior of the AIAO order parameter define by $m_{n}=(\sum^{4}_{i=1}S^{i}\cdot d^{i})/N$ ( with $d^{i}$ being denoted the four local cubic [111] directions and $N=4L^3$  the total number of spins) is plotted for $\frac{\Delta V}{V_0}=-0.2, 0.0, 0.15$ in the lattices  with the linear size $L=10$. 
The AIAO ordering vanishes for $\frac{\Delta V}{V_0}=0.2$, however, we find a phase transition for this case and we will later discuss on its detail.  
The transition temperature $T_C$ decreases by increasing the volume from $\sim 50$K for $\frac{\Delta V}{V_0}=-0.2$ to $\sim 4$K for $\frac{\Delta V}{V_0}=0.2$ (see Tab.\ref{tab1}). The  values of $T_C$ are estimated from location of the peaks in the specific heat  exhibited in Fig.\ref{fig:c}.  The last column in Table.~\ref{tab1} shows the  values of Curie-Weiss  temperature ($\Theta_{{\rm CW}}$) estimated by the linear extrapolation of the inverse susceptibility in the temperature interval $300$K to $400$K (Fig.\ref{fig:kappa}). Interestingly, the dependence of transition temperature to the ratio of the DM coupling to $J_1$ $(D/J_1)$ is found to be linear as shown in Fig.~\ref{fig:T_C}.  
\begin{table}[tp]
\centering
\caption{Critical and Curie-Weiss temperatures for different fractional volume changes.}
\begin{tabular}{cccc}
 \hline
 $\frac{\Delta V}{V_{0}}$ & $T_{C}({\rm K})$  & $\Theta_{CW}({\rm K})$  & \\
 \hline
 0.20& $\sim 4$ &$-725$ \\

 0.15&$\sim 7$ &$-814$ \\

 0.10& $\sim 11$&$-881$\\

 0.05&$\sim16$&$-912$\\

 0.00&$\sim21$&$-973$\\

 -0.05&$\sim26$&$-1082$\\

 -0.10&$\sim 33$&$-1136$\\

 -0.15&$\sim 41$&$-1194$\\

 -0.20&$\sim 50$&$-1221$\\ 
 \hline
 \label{tab1}
 \end{tabular}
\end{table}
\begin{figure}[tb]
    \centering
    \includegraphics[width=0.92\columnwidth]{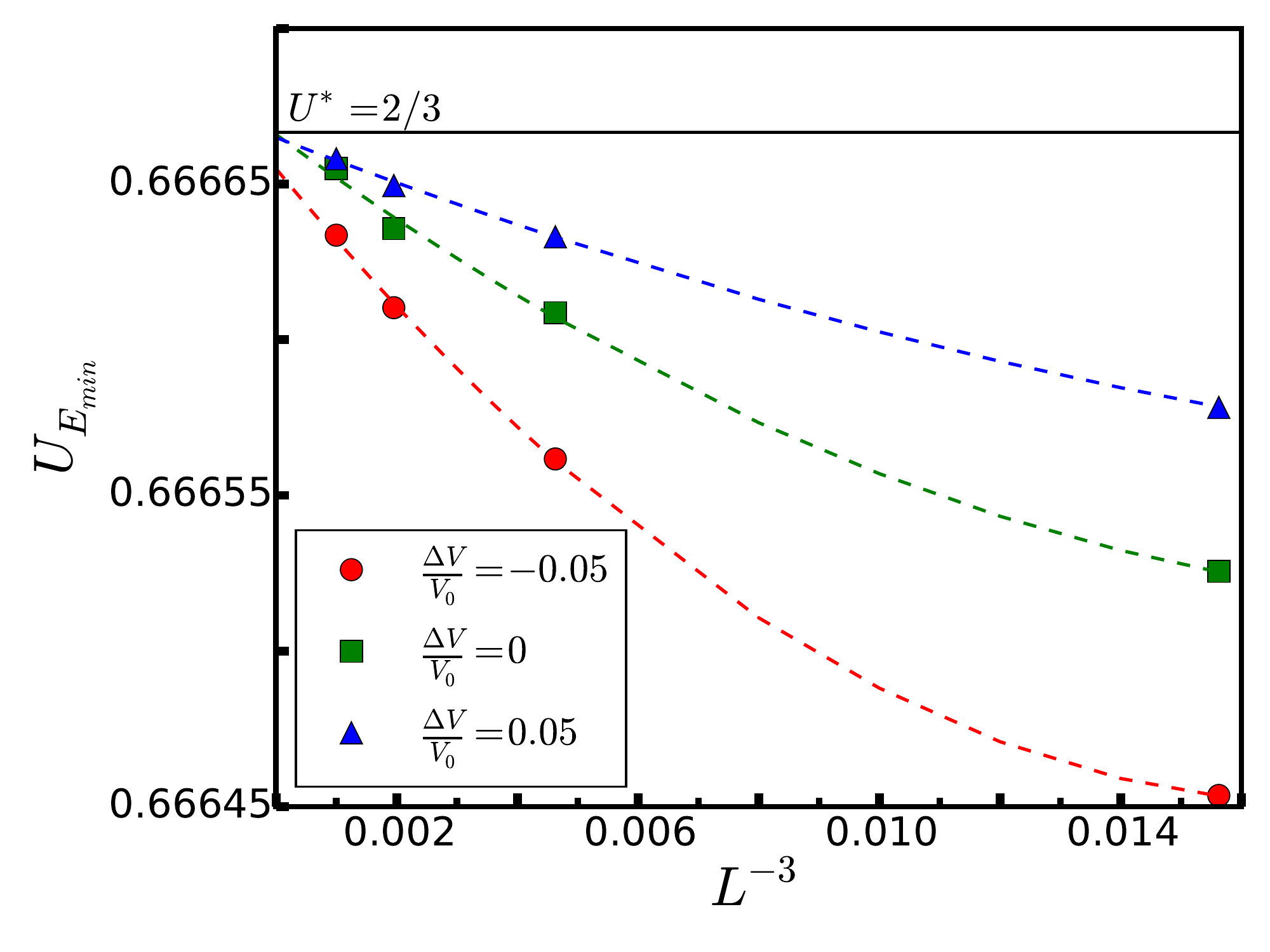} 
    \caption{(Color online) The scaling behavior of the minimum of  Binder forth energy cumulant $U_{{\rm min}}$ versus $L^{-3}$ for $\frac{\Delta V}{V_0}=-0.05, 0.0, 0.05$. The dashed lines are obtained by fitting Eq.\ref{u-scaling} to the data.}
    \label{fig:binder}
\end{figure}
\begin{figure}[bp]
    \centering
    \includegraphics[width=0.77\columnwidth]{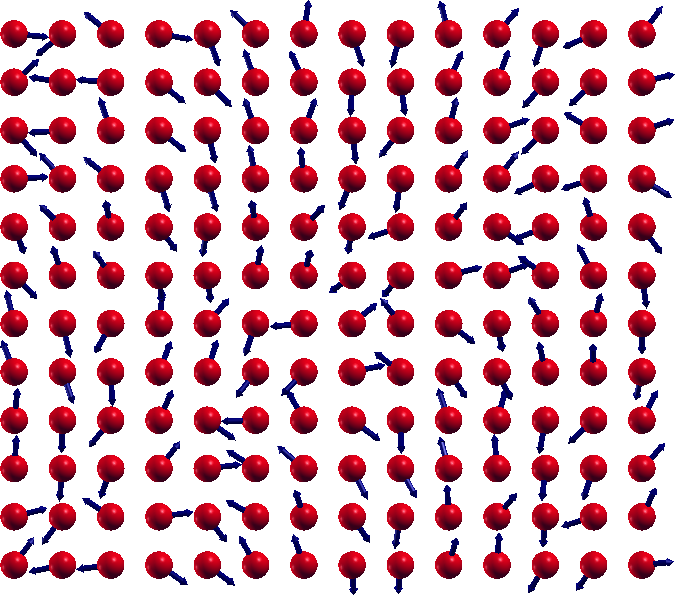}
    \caption{(Color online) A two dimensional of spin snapshot for $\frac{\Delta V}{V_{0}}=0.2$ obtained by MC simulation at $T=0.5$K for a lattice of linear size $L=10$. }
    \label{fig:spin-snapshot}
\end{figure}

\begin{figure*} [tb]
    \centering
      \includegraphics[width=0.31\textwidth]{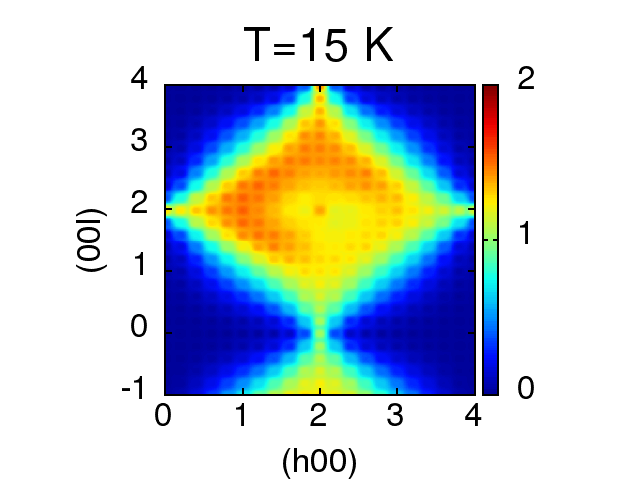}
       \includegraphics[width=0.31\textwidth]{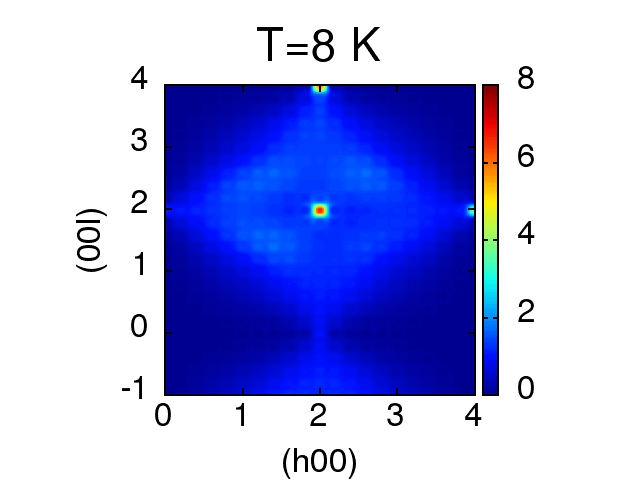}
        \includegraphics[width=0.31\textwidth]{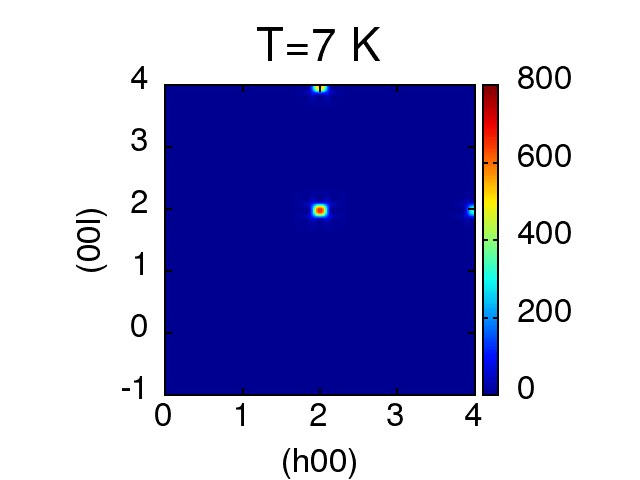}
         \includegraphics[width=0.31\textwidth]{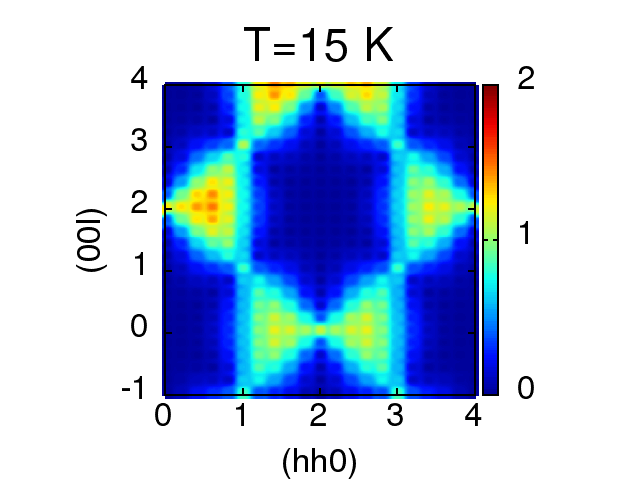}
          \includegraphics[width=0.31\textwidth]{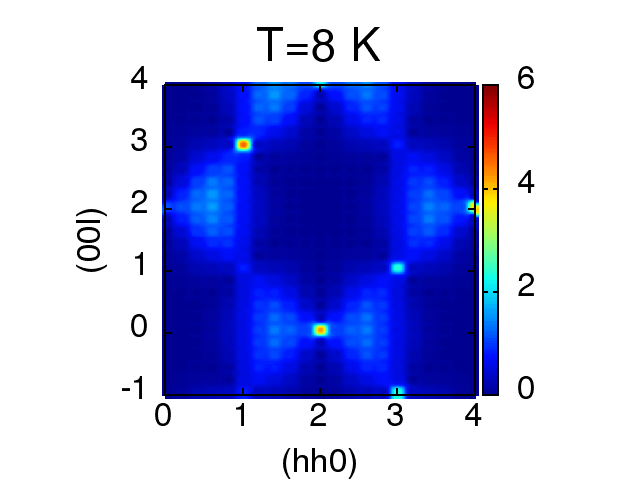}
           \includegraphics[width=0.31\textwidth]{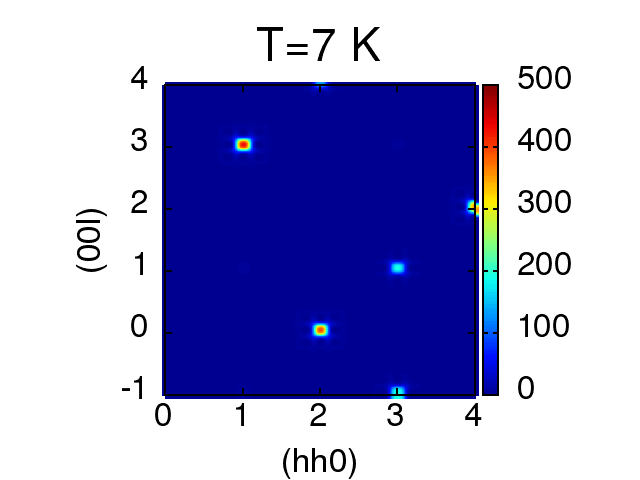}
      \caption {(Color online) The density plot of elastic neutron scattering structure function obtained by MC simulation for
       $\frac{\Delta V}{V_0}=0.15$ at $T=15, 8, 7$K in a lattice of linear size $L=10$ in  (\textbf{top}) : (h0l)  planes  and (\textbf{Bottom}): (hhl) planes.}
    \label{fig:SF0.15}
 \end{figure*}
 \begin{figure*}[tb] 
    \centering
      \includegraphics[width=0.31\textwidth]{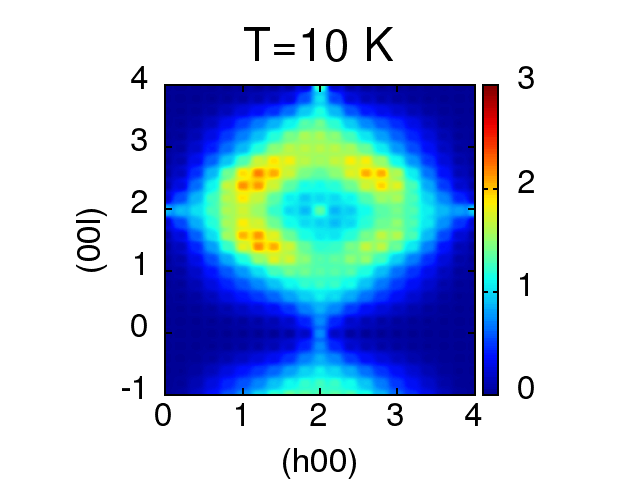}
       \includegraphics[width=0.31\textwidth]{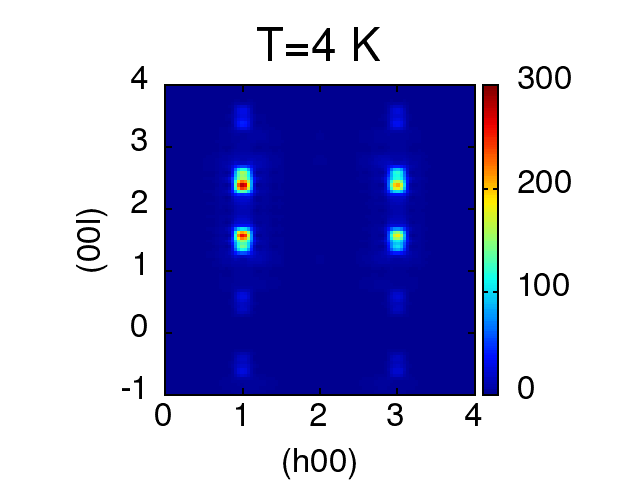}
          \includegraphics[width=0.31\textwidth]{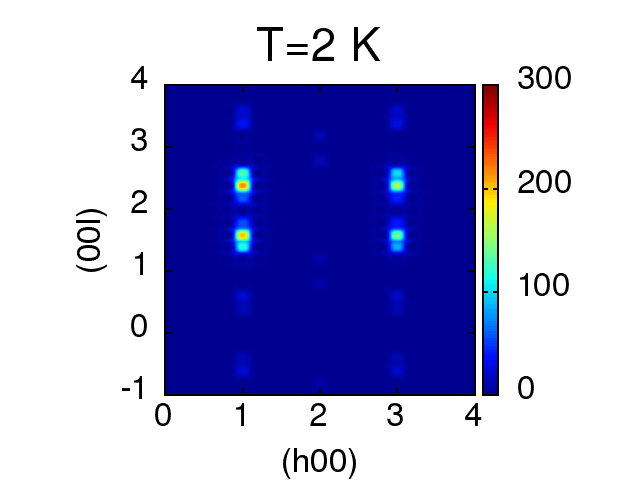}
           \includegraphics[width=0.31\textwidth]{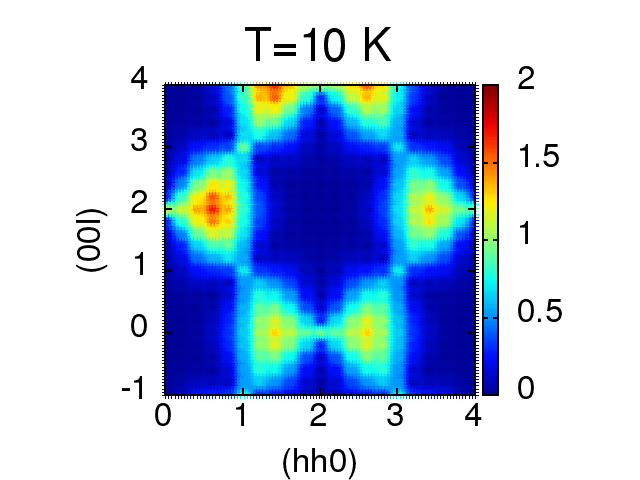}
            \includegraphics[width=0.31\textwidth]{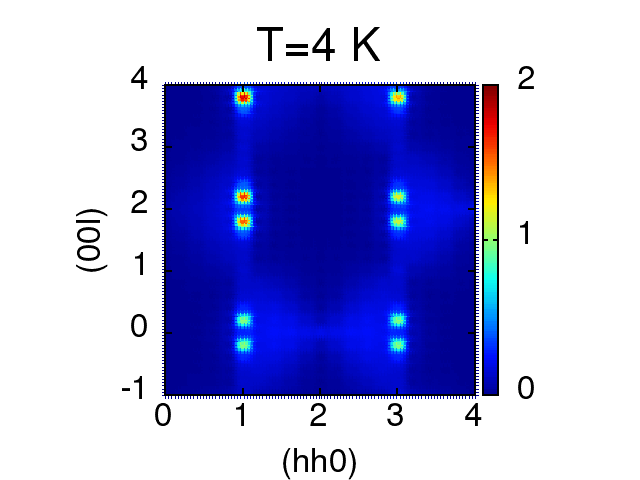}
             \includegraphics[width=0.31\textwidth]{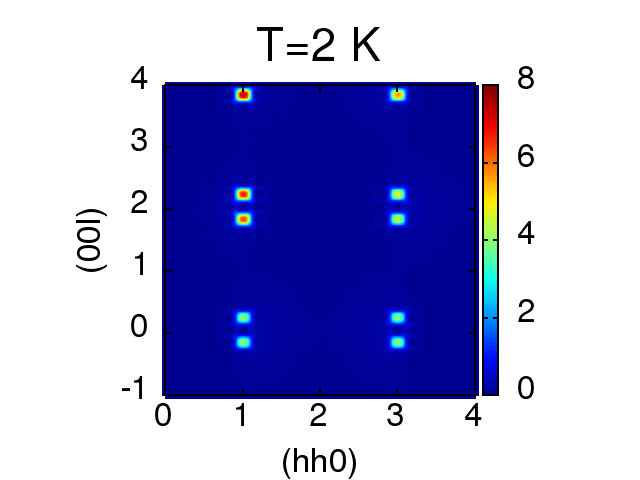}
    \caption {(Color online) The density plot of elastic neutron scattering structure function obtained by MC simulation for
       $\frac{\Delta V}{V_0}=0.2$ at $T=10, 4, 2$K in a lattice of linear size $L=10$ in  (\textbf{top}) : (h0l)  planes  and (\textbf{Bottom}):(hhl) planes.}
  \label{fig:SF0.2}
\end{figure*}
The order of transition is determined by the Binder forth energy cumulant given by 
\begin{equation}
U(T)= 1-\frac{1}{3}\frac{\langle E^4\rangle}{\langle E^2\rangle^2},
\end{equation}
in which $E$ is the total energy calculated by MC. For each lattice size, $U(T)$ shows a minimum at the transition temperature depending on the size of the lattice. The value of these minima in three dimensions obeys the following scaling behavior~\cite{binder1981}

\begin{equation}
\label{u-scaling}
U_{\rm min}(L)=U^{*}+AL^{-3}+BL^{-6}+O(L^{-9}),
\end{equation}
in which $U^{*}$ is the asymptotic value of the minimum of $U(T)$ in the thermodynamic limit. It has been proved~\cite{binder1981} that $U^{*}=2/3$ for the continuous transitions, while  for the first order transitions $U^{*} < 2/3$. 
Fig.~\ref{fig:binder} illustrates the scaling of  Binder cumulant minima versus $L^{-3}$ for $\frac{\Delta V}{V_0}=-0.05,0.0, 0.05$.
These results indicate that the order transition changes from first to second  at $\frac{\Delta V}{V_0}\sim 0.0$. This result suggests that pyr-FeF$_3$ at the ambient conditions locates in the vicinity of  a tricritical point.

Now, we discuss the case of $\frac{\Delta V}{V_0}=0.2$. As anticipated, at this volume the system does not exhibit the transition to the AIAO state, however the system undergoes a second order phase transition at $T\sim 4$K.  A  two dimensional of spin snapshot at $T=0.5$K  is illustrated  in Fig.~\ref{fig:spin-snapshot}, showing that the magnetic ground state in this case is spin modulated.   
To gain more insight into the magnetic ground states of the system we calculate the elastic neutron scattering structure function defined by 
\begin{equation}
\begin{split}
S(\mathbf q)=&\sum_{i;j}\langle (\mathbf S_{i}-\frac{\mathbf S_{i}\boldsymbol{\cdot}\mathbf q}{\mathbf q\boldsymbol{\cdot}\mathbf q}\mathbf q) \boldsymbol{\cdot} 
(\mathbf S_{j}-\frac{\mathbf S_{j}\boldsymbol{\cdot}\mathbf q}{\mathbf q\boldsymbol{\cdot}\mathbf q}\mathbf q) \rangle  \\ 
&\exp[i\mathbf q \boldsymbol{\cdot}(\mathbf R_{i}-\mathbf R_{j})].
\end{split}
\end{equation}

The density plots of $S(\mathbf q)$ for the fractional volume changes $0.15$ and $0.2$ are shown in Figs.~\ref{fig:SF0.15} and \ref{fig:SF0.2}, respectively. Both figures show the pinch point structure for $T>T_C$ which  is a peculiarity of the Coulomb phase in the Hiesenberg antiferromagnets in the pyrochlore lattice.  For $\frac{\Delta V}{V_0}=0.15$, at $T<T_C$  Bragg peaks appears at some pinch points, i.e. $(202), (220), (113)$  which are correspondent to the AIAO spin ordering~\cite{sadeghi2015}. However, for $\frac{\Delta V}{V_0}=0.2$, the Bragg peaks grow at some non-integer wave vectors, which indicates the transition to a spin modulated state. 
\subsection{critical exponents}
\begin{figure*}[t]
    \centering
    \includegraphics[width=0.95\columnwidth]{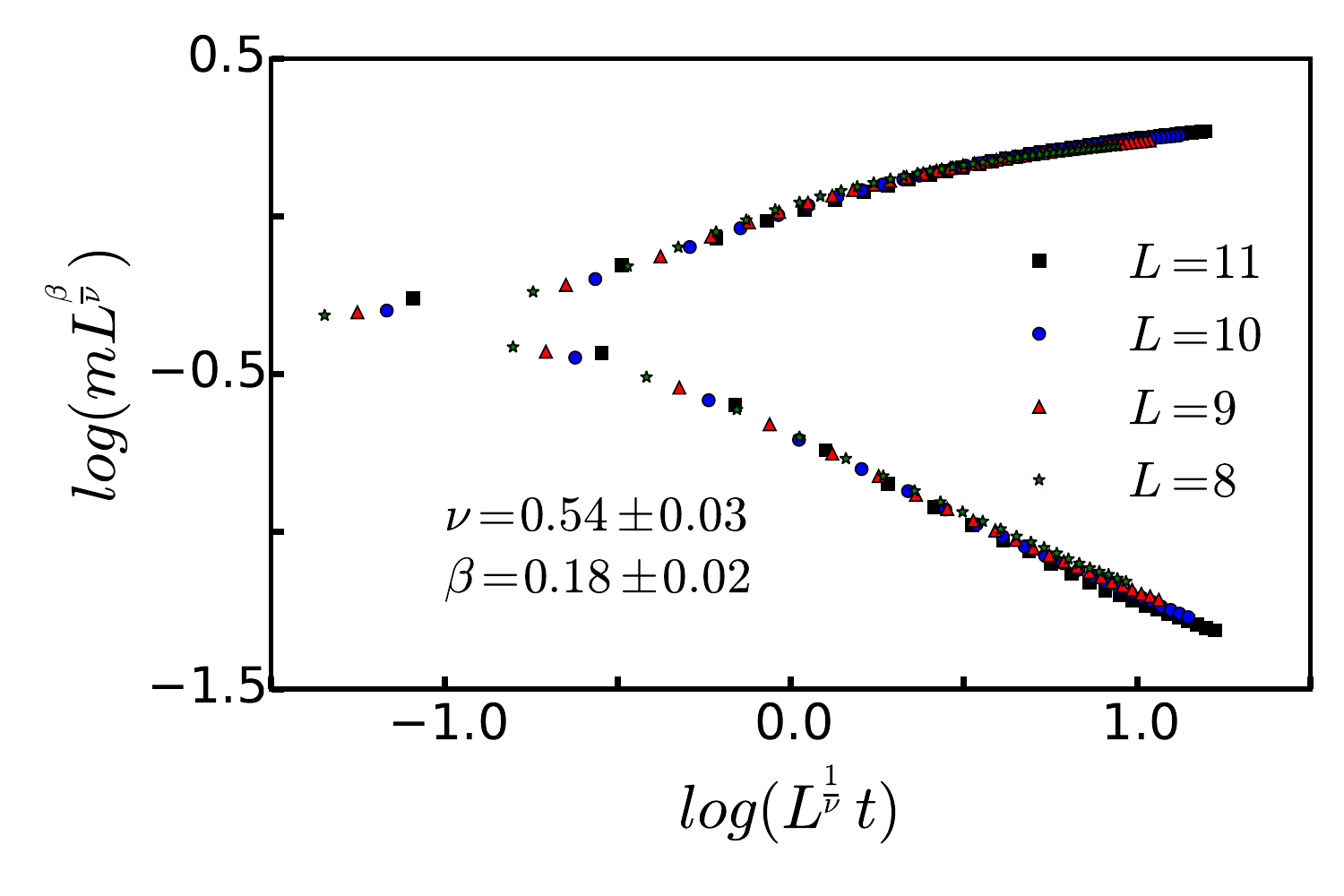}
    \includegraphics[width=0.95\columnwidth]{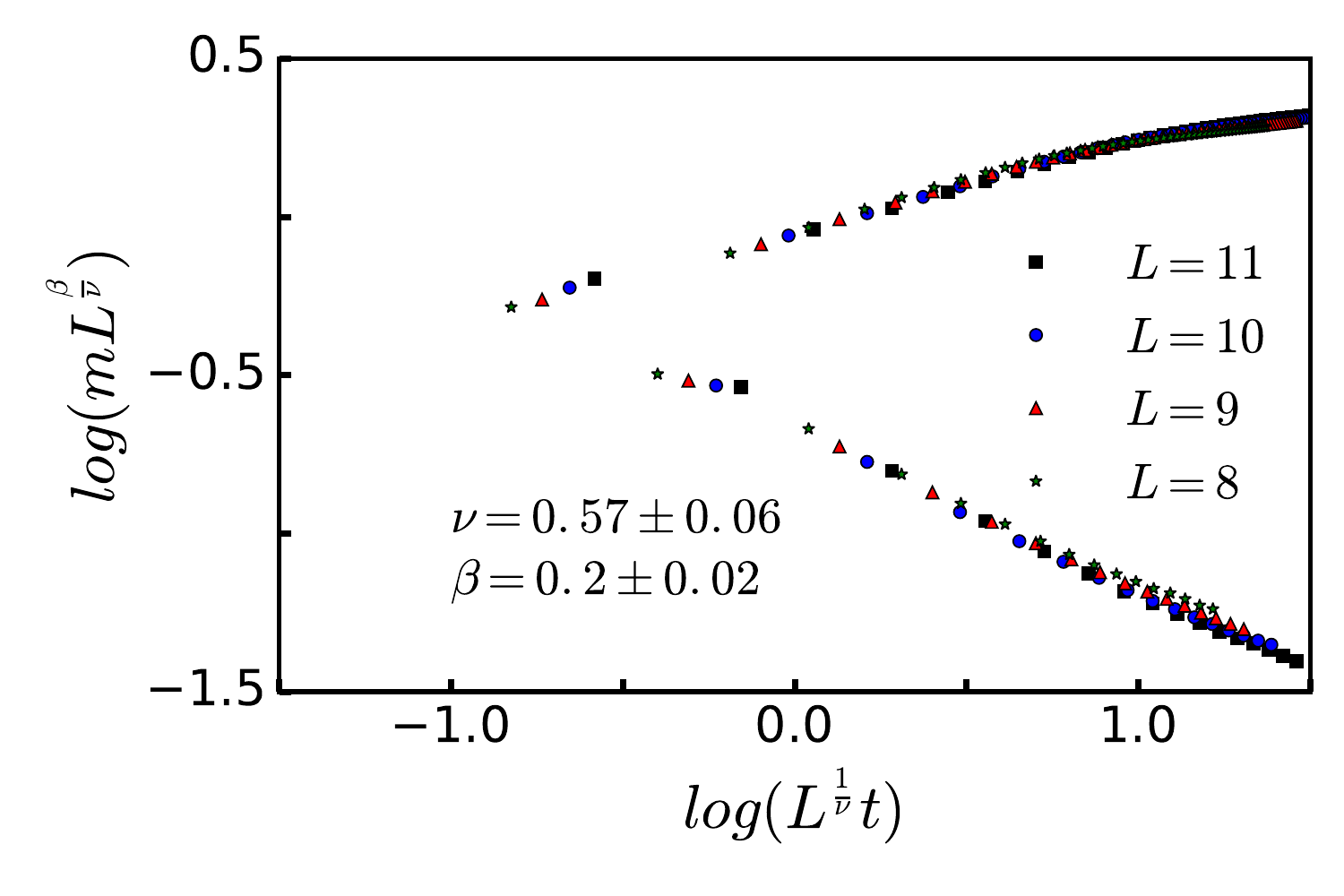}
    \caption{(Color online) Finite-size scaling of the order parameter for ({\bf Left})$\frac{\Delta V}{V_0}=0$ and ({\bf Right}) $\frac{\Delta V}{V_0}=0.15$.}
    \label{fig:beta}
\end{figure*}

We use the  finite-size scaling theory\cite{Fisher1972,Landau1976} to find the critical exponents for the second order transitions. The singular parts of the  thermodynamic quantities such as, order parameter, AIAO susceptibility and the specific heat are given by
\begin{eqnarray}
m_{n}(t, L)&\sim& L^{-{\beta}/{\nu}}{\cal F}(tL^{1/\nu}) \\
\chi(t,L)      &\sim& L^{\gamma/\nu}{\cal K}(tL^{1/\nu}) \\
c(t,L)          &\sim& L^{\alpha/\nu}{\cal C}(tL^{1/\nu}),
\end{eqnarray}
where $t=\frac{T_{C}-T}{T_{C}}$ denotes the reduced temperature and $L$ is linear lattice size.
The relation between these three exponents are given by the Rushbrooke law~\cite{KADANOFF1967} as $\alpha+2\beta+\gamma=2$.
Finite-size scaling of the AIAO order parameter for the fractional volume changes $0.0$ and $0.15$ are shown in Fig.~\ref{fig:beta}.  This figure clearly shows the data collapse of the different lattice sizes ($L=8,9,10,11$) by choosing  $\beta=0.18\pm0.02$ and $\nu=0.54\pm0.03$ for $\frac{\Delta V}{V}=0$ and $\beta=0.2\pm0.02$ and  {$\nu=0.57\pm0.06$} for $\frac{\Delta V}{V}=0.15$. 
\begin{figure*} [t]
    \centering
    \subfigure{%
      \label{fig:critical_exponent:a}
      \includegraphics[width=0.39\textwidth]{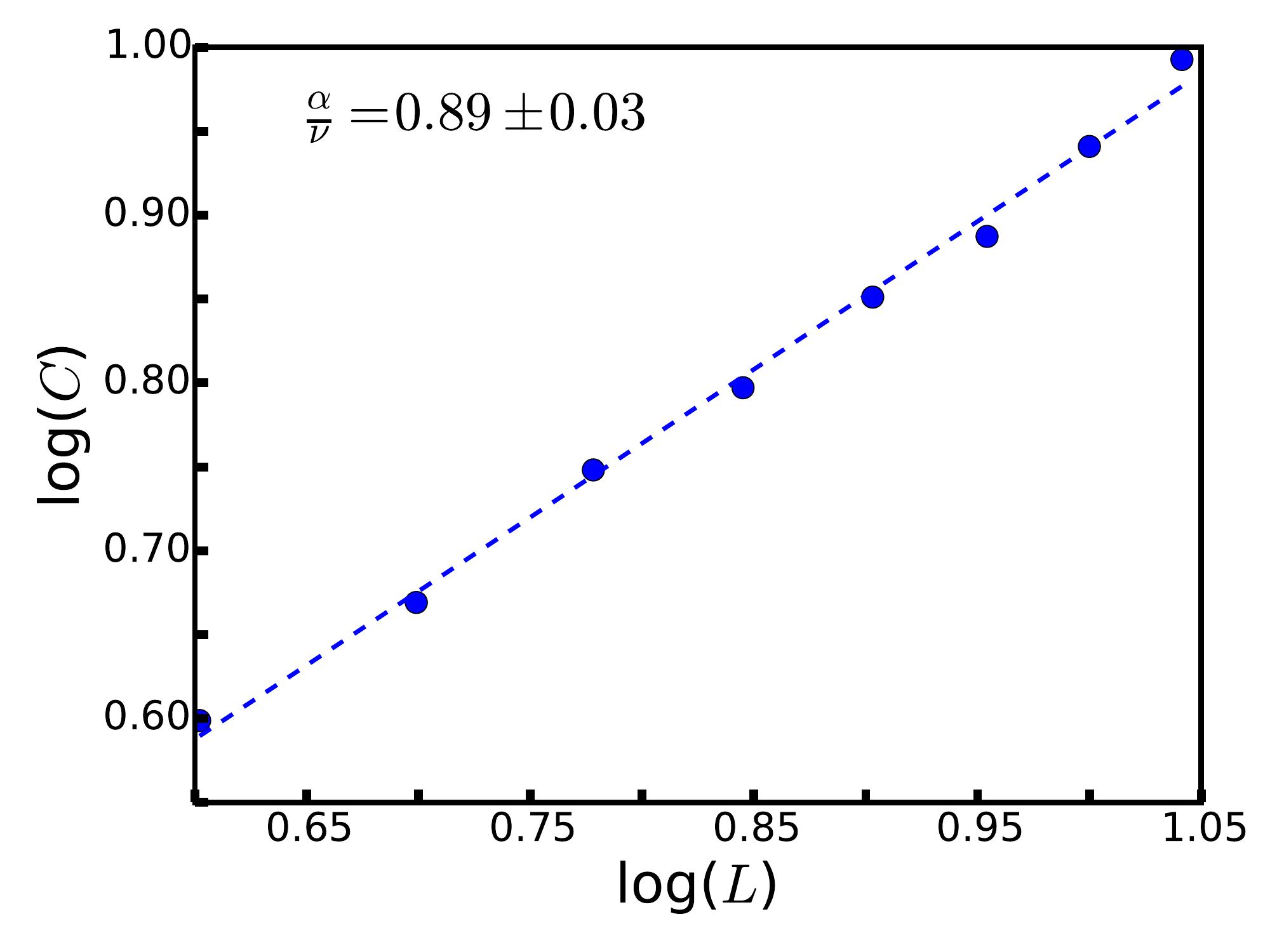}}
     \subfigure{%
       \label{fig:critical_exponent:b}
       \includegraphics[width=0.39\textwidth]{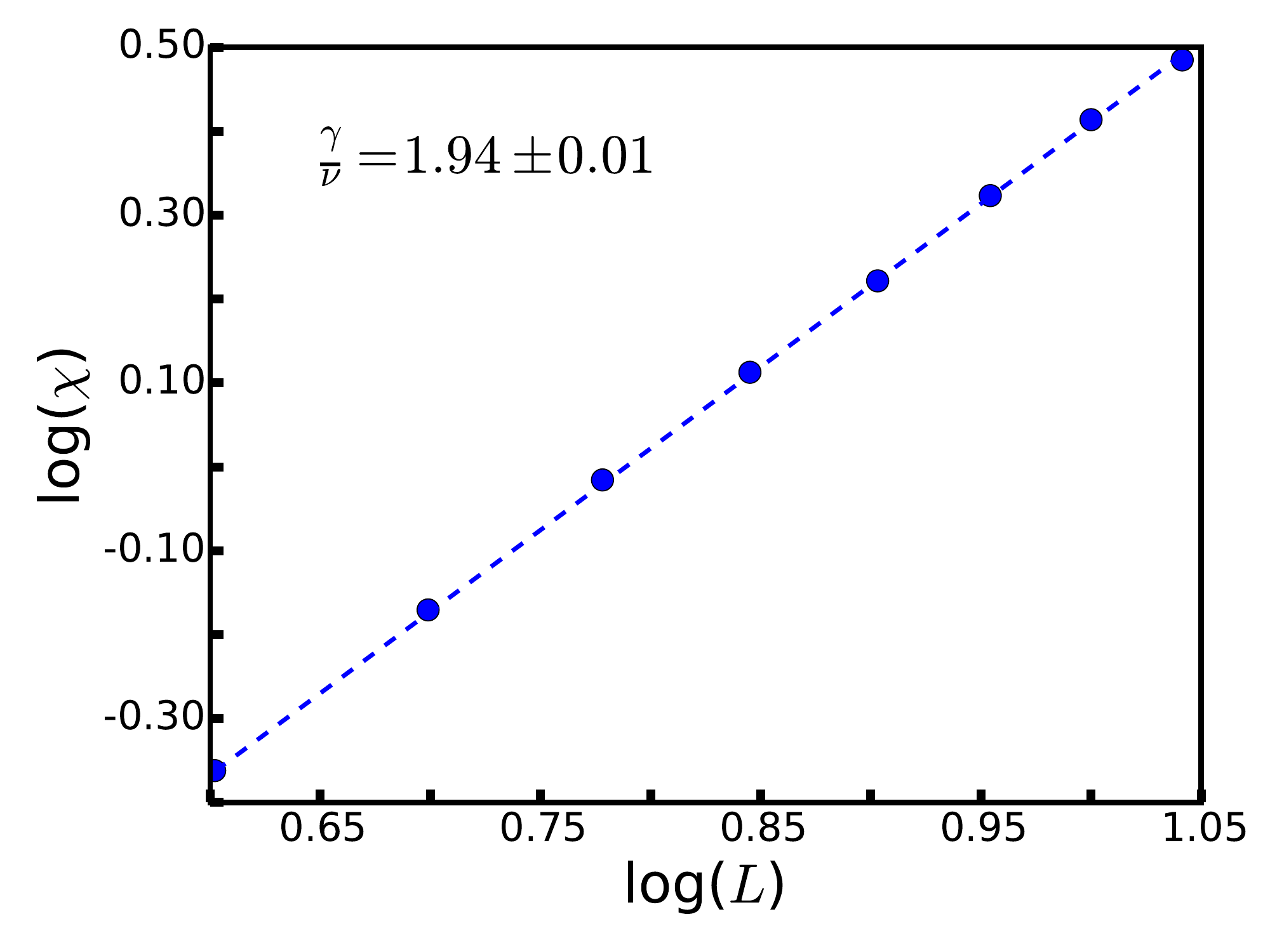}}
           \subfigure{%
      \label{fig:critical_exponent:a_1}
      \includegraphics[width=0.39\textwidth]{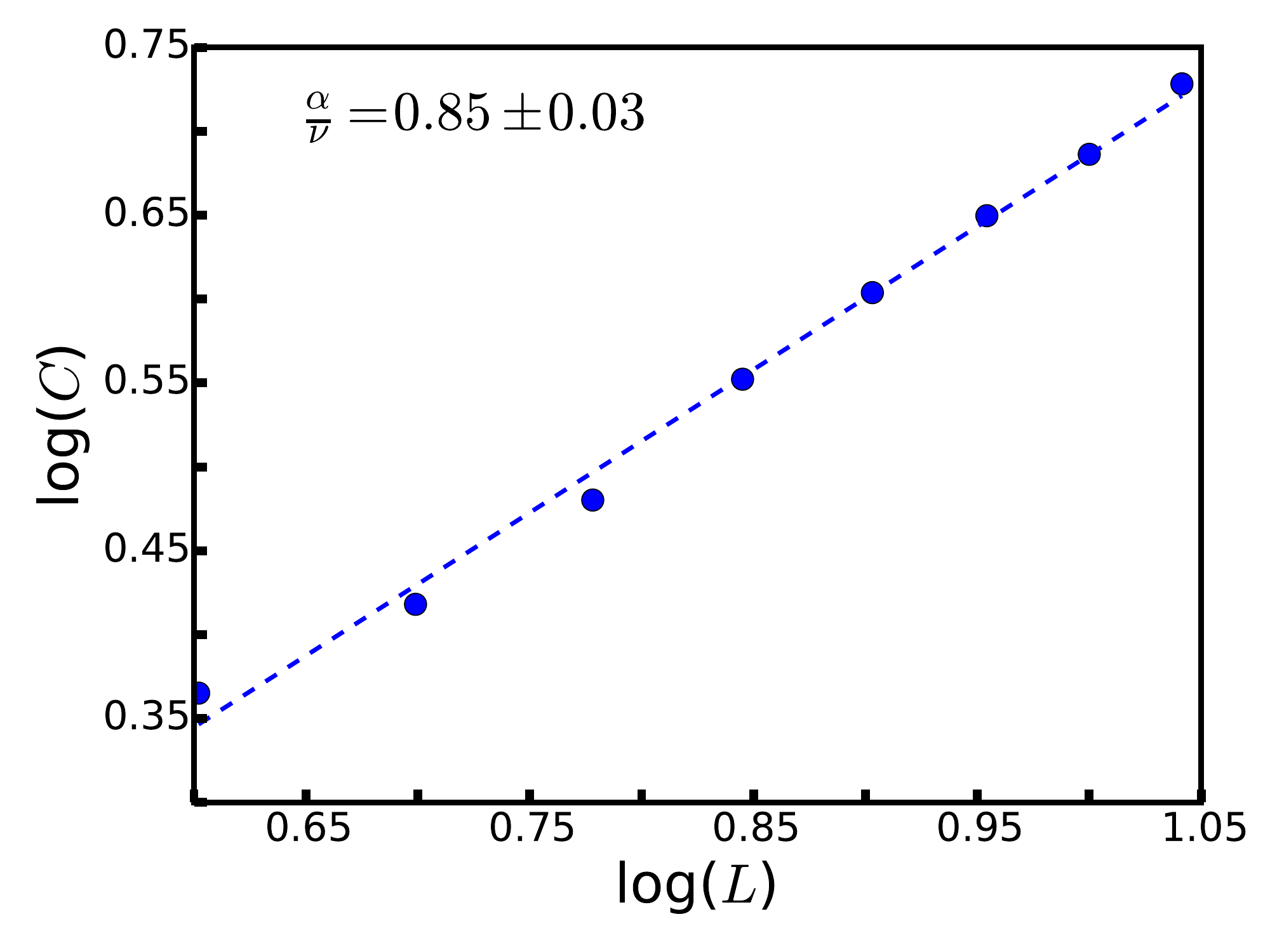}}
     \subfigure{%
       \label{fig:critical_exponent:b_1}
       \includegraphics[width=0.39\textwidth]{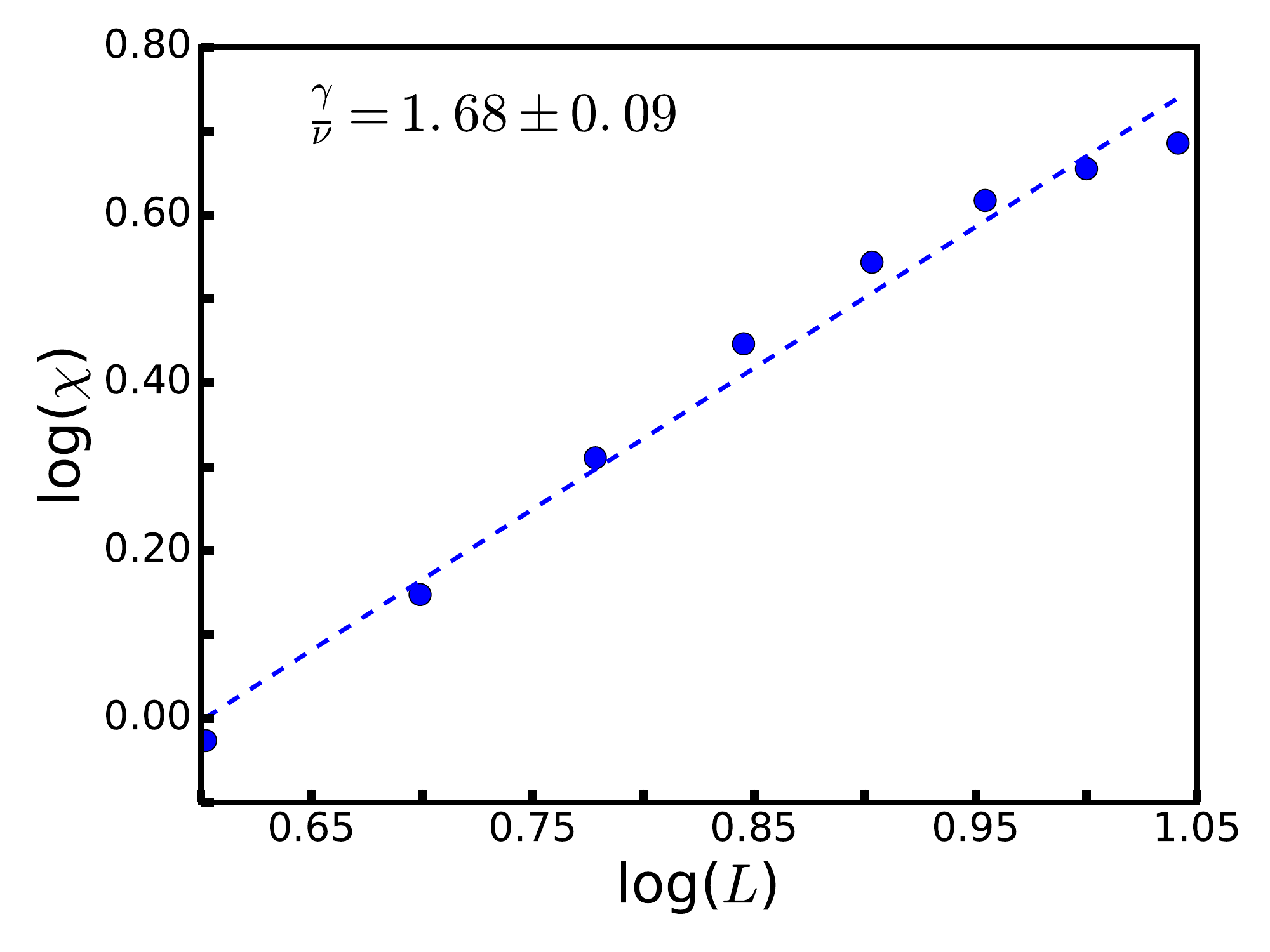}}
    \label{fig:alpha_gamma_015}
       \caption {(Color online) Logarithm of the peaks of specific heat and AIAO susceptibility versus log of the linear size of the lattice ($L=4,5,6,7,8,9,10,11$)  for ({\bf Top})$\frac{\Delta V}{V_0}=0$ and ({\bf Bottom}) $\frac{\Delta V}{V_0}=0.15$.}
    \label{fig:alpha_gamma}
\end{figure*}

To obtain the critical exponents $\alpha$ and $\gamma$, we plot the peaks of the specific heat and AIAO susceptibility versus lattice size in log-log scale (see Fig.~\ref{fig:alpha_gamma}).  The best linear fits to these data give rise to {$\alpha/\nu=0.89\pm0.03$} and $\gamma/\nu=1.94\pm0.01$ for $\frac{\Delta V}{V_{0}}=0.0$ ({$\alpha=0.48\pm 0.05$}, $\gamma=1.05\pm0.06$) and {$\alpha/\nu=0.85\pm0.03$} and {$\gamma/\nu=1.68\pm0.09$} for $\frac{\Delta V}{V_{0}}=0.15$ ({$\alpha=0.48\pm 0.06$}, {$\gamma=0.96\pm0.1$}) . It can be easily checked that  the calculated exponents $\alpha,\beta $ and $\gamma$ satisfy the Rushbrooke relation in the statistical errors
({$\alpha+2\beta+\gamma=1.92+\pm 0.15$} for $\frac{\Delta V}{V_{0}}=0.0$ and {$\alpha+2\beta+\gamma=1.85+\pm 0.2$} for $\frac{\Delta V}{V_{0}}=0.15$), moreover, both transitions are in the same universality class. 
While the exponents $\alpha$ and $\gamma$ are close to the tricritical values ($\alpha=1/2, \gamma=1$), the exponent $\beta$ deviates from the tricritical  value $1/4$ which could be due to the closeness of the Lifshitz and tricirtical   points in the parameter space of the Hamiltonian.  Interestingly, MC results for the Lifshitz point in the antiferromagnetic next to nearest neighbor Ising (ANNNI) model, have given $\beta\sim 0.2$~\cite{SELKE1988,Kaski1985}. 

In summery, using DFT calculations we found a spin Hamiltonian for pyr-FeF$_{3}$ with different volumes, including a set of coupling constants, i.e.  the AF Heisenberg exchange up to the third neighbor, the nearest neighbor bi-quadratic and direct DM interaction. The variation of the coupling constants as the function of volume change are in a way that the spin system shows a first order transition to AIAO for negative volume change with respect to the experimental structure. For the positive volume change we observed a second order  transition to AIAO state up to fractional volume change $0.15$, hence suggesting that the spin Hamiltonian corresponding to the experimental structure locates close to a tricritical point.  
At larger volume changes  the system undergoes a transition to a non-uniform spin modulated state. The reason for not having  transition to AIAO state for $\frac{\Delta V}{V_0}>0.15$ is the sign change of the bi-quadratic coupling ($B$) from positive to negative at  $\frac{\Delta V}{V_0}\sim 0.15$. In the case that both $B$ and the DM coupling are positive, they cooperate to stabilize the AIAO state at low temperatures~\cite{sadeghi2015}. However, when $B$ is negative, this term encourages  the collinear state while the DM interaction favors the AIAO, hence the competition between these two gives rise to a modular state. As the conclusion the spin system at 
$\frac{\Delta V}{V_0}=0.15$ is located at the vicinity of  a Lifshitz point. Therefore the deviation of the order parameter exponent $\beta$ from the tricritical value can be understood as the crossover from the tricritical point to a nearby lifshitz point. 
\begin{acknowledgments}
N.R and H.A acknowledge the support of the National Elites  Foundation  and  Iran  National  Science  Foundation (INSF).  
We thank Michel J.P. Gingras and Seyed Javad Hashemifar for useful discussions.
M. Amirabbasi thanks Hojjat Gholizadeh for his help in technical details.
\end{acknowledgments}
\bibliographystyle{apsrev4-1}
\bibliography{bib}
\end{document}